# Modeling Vanilla Option prices: A simulation study by an implicit method

Snehanshu Saha [*,1]     Swati Routh [2]     Bidisha Goswami [1]

1-Department of Computer Science & Engineering,
PES Institute of Technology-Bangalore South Campus
Bangalore-56100
snehanshusaha@pes.edu

2- Department of Physics, CPGS, Jain University
Bangalore-560011
swati.routh@jainuniversity.ac.in

1-Department of Computer Science & Engineering,
PES Institute of Technology-Bangalore South Campus
Bangalore-56100
bidishagoswami@pes.edu

*-communicating author

## ABSTRACT:

Option contracts can be valued by using the Black-Scholes equation, a partial differential equation with initial conditions. An exact solution for European style options is known. The computation time and the error need to be minimized simultaneously. In this paper, the authors have solved the Black-Scholes equation by employing a reasonably accurate implicit method. Options with known analytic solutions have been evaluated. Furthermore, an overall second order accurate space and time discretization has been accomplished in this paper.

## INTRODUCTION

Suppose a person possessing Microsoft shares, calls in today with the following offer: in 3 months' time, you have the option to purchase Microsoft shares from him for $50 per share.

The key point is that an individual has the option to buy the shares. Three months from now, he/she will check the market price and decide whether to exercise that option. This deal has no downside - three months from now he/she either makes a profit or walks away unscathed, On the other hand, the seller has no potential gain and an unlimited potential loss[*]. To compensate, there will be a cost for the individual to enter into the option contract. He/she must pay the seller some money up front.

The option valuation problem [1, 7] is thus to compute a fair value for the option. More precisely, it is to compute a fair value at which the option may be bought and sold in an open market. This option is described as a European Call or vanilla option.

## RELATED WORK

Digital options are exotic options with a discontinuous initial condition [8]. Whereas Crank-Nicolson shows oscillations in hedge parameter [8], BDF4 [5, 8] is accurate in all hedge parameters. With the transformed fourth order accurate scheme, the coordinate S = E must be exactly between two grid points for fourth order convergence. Then, also for digital options, a small error is obtained with a few grid points. The implied volatility modeling is also used to value an option with given option and asset contracts. The solution is typically obtained in fewer than 10 iterations by applying the inverse quadratic interpolation [9]. The error decreases almost quadratically.

Numerical experiments are necessary to determine the values of European Call/Put options and test for accuracy. The reasons behind Choosing Crank-Nicholson Scheme are *s*econd order accuracy and unconditional stability. The methods mentioned above don't posses unconditional stability and is the strength of the proposed finite difference method [4].

Our work is organized as follows: Section III briefly discusses the governing differential equation, followed by section IV where the numerical framework is elucidated. Section V contains a couple of illustrations involving the Put/Call [7] options. Notations and terminologies are defined in appendices A and B respectively. Appendix C contains a brief discussion of the analytical solution to the Black-Scholes partial differential equation for the general readership.

## THE BLACK-SCHOLES EQUATION

### Assumptions:

The fundamental assumption made about the random movement of asset prices for a more flexible hypothesis.
- We assume that present price is a full reflection of the past history experienced. And does not contain any more information;
- We assume that we would obtain immediate response for updated information about an asset.

### *Model*:

Asset price modeling identifies modeling the arrival of new information which affects the price of an asset. Absolute change in an asset is a relative measurement of the change in price. This is a better indicator of its size aspect than absolute measures.

Let's consider that at time $t$, the asset price is $S$. Let us consider a small subsequent time interval $dt$, during which $S$ changes to $S + dS$. Now, there is deterministic and anticipated return represented by $\mu dt$, where $\mu$ is known as drift. The other factor affecting the asset price is the random change in the asset price in response to external effects. It is represented by a random sample drawn from a normal distribution with mean zero denoted as $\sigma dX$, Putting the above two factors together we obtain a stochastic differential equation,

$$\frac{dS}{S} = \sigma dX + \mu dt, \qquad (1)$$

which is the mathematical representation for generating asset prices [1, 3].

The equation (1) gives interesting and important information concerning the behavior of $S$ in a probabilistic sense. At time $t = t'$, suppose the price is $S'$, and then $S'$ will be distributed about $S_0$ with a probability density function. The future asset price $S'$ is thus most likely to be close to $S_0$ and less likely to be far away. Thus, the equation generates time series – each time the series is restarted a different path results. Each path is called a **realization** of the random walk.

The **Black-Scholes Partial Differential Equation,** the derivation of which is well-known, is given below

$$\frac{\partial V}{\partial t} + \frac{1}{2}\sigma^2 S^2 \frac{\partial^2 V}{\partial S^2} + rS\frac{\partial V}{\partial S} - rV = 0 \qquad (2)$$

If $S > E$ at expiry, then in financial sense, for call option, there will be profit $S - E$, handing over an amount $E$, to obtain an asset worth, while, if $S < E$, then there will be loss of $E - S$. Thus the value of the call option at expiry can be written as

$$C(S,T) = \max(S - E, 0) \qquad (3)$$

As the time tends to expiry date the value of call option approaches (6), it is known as **pay-off function** for *European Call Option*. This is known as the final condition of PDE (12).

Now, from (1), if $S = 0$ then $dS = 0$ which means pay-off is also zero. Thus, the call option is worthless on $S = 0$ even if there is long time to expiry. Hence, we have

$$C(0,t) = 0 \qquad (4)$$

And if $S \to \infty$ i.e. the asset price increases without bound it becomes ever more likely that the option will be exercised and the magnitude of the exercise price becomes less and less important. Thus as $S \to \infty$ the value of the option becomes that of the asset and so

$$C(S,t) \sim S \text{ as } S \to \infty \qquad (5)$$

*Thus, the Black Scholes equation and boundary condition for a **European Call Option** is given by equations (2)-(5).*

For *European Put Option*, the final condition is the payoff

$$P(S,T) = \max(E - S, 0) \qquad (6)$$

And the initial condition $P(0,t)$ is determined by calculating the present value of an amount $E$ received at time $T$. For time-dependent interest rate, the boundary condition at $S = 0$ is

$$P(0,t) = Ee^{-\int_t^T r(T)dT} \qquad (7)$$

And,

$$P(S,T) \to 0 \text{ as } S \to \infty \qquad (8)$$

*Thus, the Black Scholes equation and boundary conditions for a **European Put Option** are given by equations (2), (6)-(8).*

The analytical solution to the BSDE is not the focal point of the paper but given below, nonetheless as

$$C(S,t) = SN(d_1) - Ee^{-r(T-t)}N(d_2) \qquad (9)$$

where

$$d_1 = \frac{\log(S/E) + \left(r + \frac{1}{2}\sigma^2\right)(T-t)}{\sigma(\sqrt{T-t})}$$

$$d_2 = \frac{\log(S/E) + \left(r - \frac{1}{2}\sigma^2\right)(T-t)}{\sigma(\sqrt{T-t})}$$

The corresponding calculation for a European Call Option follows similar lines.

## NUMARICAL SOLUTION

### A. Finite Difference Methods

The finite difference method essentially converts the ODE [2] into a coupled set of algebraic equations, with one balance equation for each finite volume/node in the system. The general technique is to replace the continuous derivative within the ODE with finite-difference approximation on a grid of mesh points that spans the domain of interest. A very interesting implicit method, The Runge-Kutta method has been detailed in [5].

A key step in the finite difference method is to replace the continuous derivative in the original ODE with appropriate approximation in terms of the dependent variable evaluated at different mesh points. Let's evaluate the continuous variable $x$ at discrete points $x_i$

$$x_i = x_0 + i\Delta x = x_0 + ih;$$
$$h = |\Delta x|$$

For the dependent variable, $f(x)$ the discrete representation becomes

$$f(x_i) \to f_i$$
$$f(x_i + h) \to f_{i+1}$$
$$f(x_i - h) \to f_{i-1}$$

Using Taylor Series for i+1 and i-1 and subtracting, we obtain

$$f_{i+1}(x+h) - f_{i-1}(x-h) \to$$

$$2f'_i(x)h + 2f'''_i(x)\frac{h^3}{3!} + \ldots$$

$$f'_i(x) = \frac{f_{i+1}(x+h) - f_{i-1}(x-h)}{2h} + O(h^2) \qquad (10)$$

where $O(h^2)$ is the truncation error of the order of $h^2$. This is the First Order Differentiation approximation.

Similarly by adding equations for i+1 and i-1, we have

$$f''_i(x) = \frac{f_{i+1}(x+h) - 2f_i(x) + f_{i-1}(x-h)}{h^2} + O(h^2) \qquad (11)$$

This is known as the Second Order Differentiation approximation.

Crank-Nicholson's Implicit Method

The heat equation in one spatial variable accompanied by boundary conditions appropriate to a certain physical phenomenon is given as

$$\frac{\partial^2}{\partial x^2} u(x,t) = \frac{\partial}{\partial t} u(x,t)$$

with initial conditions
$$u(0,t) = u(1,t) = 0$$
and boundary condition
$$u(x,0) = \sin \pi x$$

(12)

(12)

Now, substituting equation (10) and (11) in the differential equation (12), with possibly different step lengths $h$ and $k$, the central differences at $(x, t - \frac{k}{2})$ the result turns out to be

$$\frac{1}{h^2}\left[u\left(x+h, t-\frac{k}{2}\right) - 2u\left(x, t-\frac{k}{2}\right) + u\left(x-h, t-\frac{k}{2}\right)\right] = \frac{1}{k}\left[u(x,t) - u(x, t-k)\right] \quad (13)$$

The locations of the six points in the equation in the figure below –

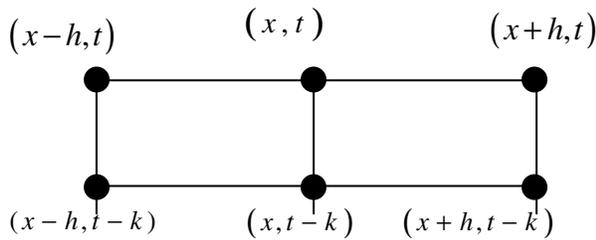

$(x-h, t) \quad (x, t) \quad (x+h, t)$

$(x-h, t-k) \quad (x, t-k) \quad (x+h, t-k)$

Since the $u$ values are known only at integer multiples of $k$, terms like $u\left(x, t - \frac{k}{2}\right)$ are replaced by the average of $u$ values at adjacent grid points, i.e.

$$u\left(x, t - \frac{k}{2}\right) \approx \frac{1}{2}\left[u(x,t) + u(x, t-k)\right]$$

So, we have

$$\frac{1}{2h^2}\begin{bmatrix} u(x+h,t) - 2u(x,t) + u(x-h,t) + \\ u(x+h,t-k) - 2u(x,t-k) + u(x-h,t-k) \end{bmatrix} = \frac{1}{k}\left[u(x,t) - u(x, t-k)\right]$$

The computational form of this equation is

$$-u(x-h,t) + 2(1+s)u(x,t) - u(x+h,t)$$
$$= u(x-h,t-k) - 2(1-s)u(x,t-k) + u(x+h,t-k) \quad (14)$$

where $s = \dfrac{h^2}{k}$

This leads to a tri-diagonal system of the form

$$\begin{bmatrix} r & -1 & 0 & 0 & . & . & . & . & . & 0 \\ -1 & r & -1 & 0 & 0 & . & . & . & . & 0 \\ 0 & -1 & r & -1 & 0 & 0 & . & . & . & 0 \\ . & . & . & . & . & . & . & . & . & . \\ . & . & . & . & . & . & . & . & . & . \\ 0 & . & . & . & . & 0 & 0 & -1 & r & -1 \\ 0 & . & . & . & . & . & 0 & 0 & -1 & r \end{bmatrix} \begin{bmatrix} u_1 \\ u_2 \\ . \\ . \\ . \\ u_{n-2} \\ u_{n-1} \end{bmatrix} = \begin{bmatrix} b_1 \\ b_2 \\ . \\ . \\ . \\ b_{n-2} \\ b_{n-1} \end{bmatrix} \quad (15)$$

with $r = 2(1+s)$ and

$$b_i = u((i-1)h, t-k) + 2(s-1)u(ih, t-k) + u((i+1)h, t-k)$$

*Numerical Stability of Crank-Nicholson's Method*:

$$\frac{F(x, t_{n+1}) - F(x, t_n)}{k} = D \frac{\partial^2 F(x, t_n)}{\partial x^2} + O(h^2)$$

If we evaluate the right-hand side partly (wholly) at the end of time step $t_{n+1}$, then C-N method involves the average of right-hand side b/w the beginning and end of time step.

$$\frac{F(x, t_{n+1}) - F(x, t_n)}{k} = \frac{D}{2} \frac{\partial^2 F(x, t_n)}{\partial x^2} + \frac{D}{2} \frac{\partial^2 F(x, t_{n+1})}{\partial x^2} + O(h^2)$$

The accuracy of Crank-Nicholson method is $O(h^2)$ as seen from above.

Using Central Difference to approximate $\frac{\partial^2}{\partial x^2}$

$$\frac{F_i^{n+1} - F_i^n}{k} = D \frac{F_{i-1}^n - 2F_i^n + F_{i-1}^n}{h^2}$$

$$F_i^{n+1} = F_i^n + C(F_{i-1}^n - 2F_i^n + F_{i-1}^n) \quad (16)$$

for $i = 1, \ldots, N$, $F_i^n = F(x_i, t_n)$, $C = D \frac{k}{(h)^2}$

Using this C-N scheme becomes

$$F_i^n + \frac{C}{2}(F_{i-1}^n - 2F_i^n + F_{i+1}^n) = \\ F_i^{n+1} - \frac{C}{2}(F_{i-1}^{n+1} - 2F_i^{n+1} + F_{i+1}^{n+1}) \quad (17)$$

Now using Von Neumann stability analysis,

$$F(x, t) = \hat{F}(t) e^{ikx}$$

Equation (16) becomes

$$\hat{F}^{n+1}e^{iks} = \hat{F}^n e^{iks}\left[1+C\left(e^{-ik\delta s}-2+e^{ik\delta s}\right)\right]$$

$$\hat{F}^{n+1} = A\hat{F}^n\ ; A = 1-2C(1-\cos n\delta x) \tag{18}$$

Using equations (16) and (17),

$$A = \frac{1-4C\sin^2\left(k\dfrac{\delta x}{2}\right)}{1+4C\sin^2\left(k\dfrac{\delta x}{2}\right)} < 1, \forall\, C$$

Since, the amplification factor is less than 1 regardless of the input values the Crank-Nicholson method is unconditionally stable.

So, we obtain a tri-diagonal matrix linking $F_i^{n+1}$ and $F_i^n$ at every step of the scheme. Thus the price we pay for the high accuracy and unconditional stability of the scheme is the necessity to invert a tri-diagonal matrix equation at each time step. But the trade-off is profitable enough!

**Implementation Details: the pseudo code**

```
% Finite Difference Methods used to solve the Black-Scholes PDE equation, a parabolic Partial Differential Equation

% Type of Option
opt = input('Call or Put Option(0 or 1): ');

% % Parameters
E=input('Agreed Exercise Price(Money): ');
r=input('Continously Compounded Interest Rate: ');
T=input('Expiry Date(in years): ');
% sig=input('Volatility of the Assest(0<=sigma<=1): ');
a = rand(5);
for i = 1:size(a)
   for j = 1:size(a)
      sig(i,j) = a(i,j);
      Nx=11;  Nt=29; L=10; {Set the parameters}

      % Call the CrankNicholsonMethod
      u2 = CrankNicholsonMethod(opt,E,sig(i,j),r,T,Nx,Nt,L);
%      u2 = CrankNicholsonMethod(opt,E,sig,r,T,Nx,Nt,L);
      surf(u2)
      hold on
   end
end;
 Assign Labels: xlabel('Asset Value');ylabel('Time');zlabel('Payoff/Exercise Value');
```

**CrankNicholsonMethod**

```
% Pricing a European option using the Crank-Nicolson method on the Black-Scholes PDE
function [U] = CrankNicholsonMethod(type,K,sigma2,r,T,N,M,Xmax)
 fdeltat = T/N; deltaX = Xmax/M; (Threshold)
% Initializing matrices according to the tri-diagonal structure
X = [];    % Exercise Price;U = [];temp = zeros(M-1,M+1);

for i = 1:M+1
Use U to compute the payoff matrix;
```

```
 % The boundary values for two different types- Put and Call
if type == 0
   for j = 2:N+1
      U(1,j)=0;
      U(M+1,j) = Xmax - K*exp(-r*((j-1)*deltat));
   end
else
   for j = 2:N+1
      U(1,j) = K;
      U(M+1,j) = 0;
   end
end
Compute  the Matrices and diagonalize;

% Execute the time loop
for j = 1:N
    matrixB = U(:,j);
    matrixC(1,1) = temp(1,1)*U(1,j+1);
    matrixC(M-1,1) = temp(M-1,M+1)*U(M+1,j+1);
    known = matrixA*matrixB - matrixC;
    U_jplus1 = zeros(M-1,1);
    U_jplus1 = inv(matrixD)*known;
    for k = 1:M-1
       U(k+1,j+1)= U_jplus1(k,1);
    end
end

% Function of the payoff
function y = payoff(type,z,K)
if type == 0
   y = max(z-K,0);   % Call Option
else y = max(K-z,0);  % Put Option
end
```

## SIMULATION RESULTS

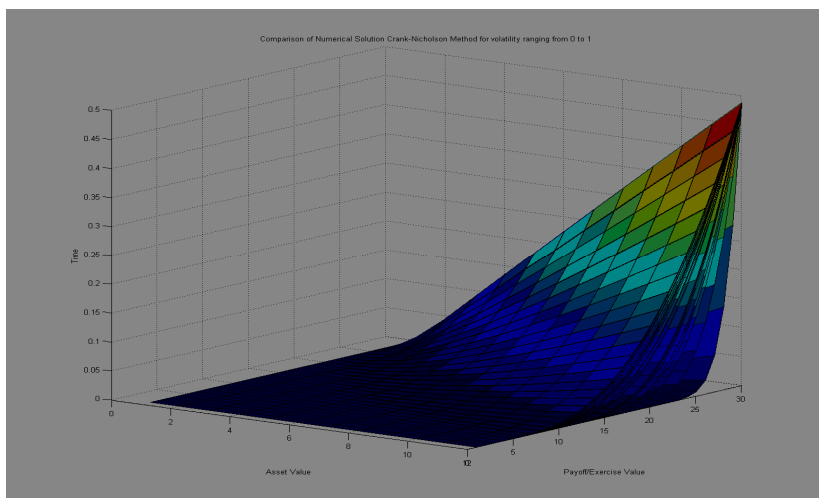

For Call Option, Agreed Price = $10, Rate of Interest = 0.1, Expiry Date = 6 months.

The Black Scholes model is obtained as a solution to a parabolic PDE (called the Black Scholes PDE) for pricing an option for an underlying asset. If the asset is volatile then pricing the option through a model is particularly helpful to determine the Payoff function. Crank Nicholson implicit scheme is more realistic among the finite difference methods in the sense that it is stable regardless of the parameters. Although it is more complicated to implement Crank Nicholson scheme, unconditional stability is too good an issue to compromise and hence used as the numerical technique to solve the PDE in our work. Future work may include looking at parallel computing methods to arrive at the result faster.

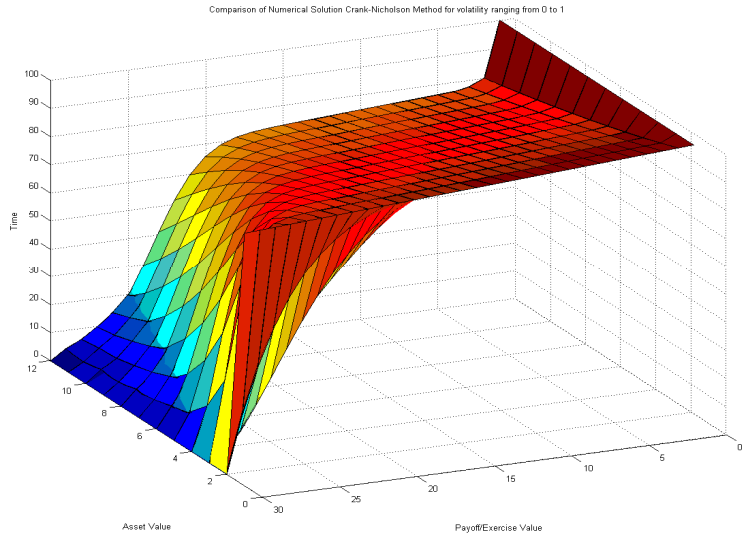

For Put Option, Agreed Price = $100, Rate of Interest = 0.25, Expiry Date = 12 months.

The simulation study results in obtaining/fixing the price of a European option with given agreed prices for Call and put options respectively. The output is a time –series representation when the volatilities are varied in the range of 0 and 1. The results show the advantages of such modeling e.g.

- ➢ The asset price follows lognormal random walk.
- ➢ The underlying asset doesn't pay any dividends during the lifespan of the option.
- ➢ There are no arbitrage possibilities.
- ➢ Trading of the underlying asset could be continuous.
- ➢ Short selling is permitted and the assets are divisible.

The volatility of the asset is well captured and the profile doesn't become unbounded in finite time i.e. stability is accomplished which is the key reason for employing Crank-Nicholson's method.

**Appendix A: Notation Table**

1.) $t$ - Time
2.) $S_0$ - Value of the asset at $t_0 = t$
3.) $S$ - Asset Price
4.) $E$ - Exercise Price
5.) $d*$ - Change in any quantity over a time interval
6.) $\mu$ - Measure of the average rate of growth of the asset price
7.) $\sigma$ - Volatility, measures the standard deviation of the returns
8.) $dX$ - Random variable, drawn from a normal distribution
9.) $\phi$ - A random variable from a standardized normal distribution
10.) $F(*)$ - Function
11.) $r$ - Risk-free interest rate
12.) $V$ - Value of an option, a function of $S$ and $t$
13.) $C$ - Call Option function
14.) $P$ - Put Option function

**Appendix B: Terminologies**

1.) **Writer**: The person **selling** the asset.

2.) **Holder**: The person **buying** the asset.

3.) **Expiry Date**: At a **prescribed time** in the future, the holder/writer of the option may purchase/sell a prescribed asset.

4.) **Underlying Asset**: The holder/writer may purchase/sell a **prescribed asset** at a prescribed time.

5.) **Exercise Amount/Strike Price**: The holder/writer purchases a prescribed asset at a prescribed time at a **prescribed amount**.

6.) **Arbitrage:** In financial terms, there are never any opportunities to make an instantaneous risk-free profit.

7.) **Risk:** Risk is commonly of two types – specific and non-specific, specific risk is the component of risk associated with a single asset, whereas non-specific risk is associated with factors affecting the whole market.

**Appendix C. Analytical Solution**

$$\frac{\partial V}{\partial t} + \frac{1}{2}\sigma^2 S^2 \frac{\partial^2 V}{\partial S^2} + rS\frac{\partial V}{\partial S} - rV = 0$$

with initial condition

$$C(S,T) = \max(S - E, 0)$$
$$C(0,t) = 0$$

and boundary condition

$$C(S,t) \sim S \text{ as } S \to \infty$$

For *European Call Option,* suppose

$$S = Ee^x, \quad t = T - \tau \Big/ \tfrac{1}{2}\sigma^2, \quad C = Ev(x,\tau) \tag{i}$$

$$V \to v \ \& \ S \to x$$

Then the **Black-Scholes Partial Differential Equation** results in

$$\frac{\partial v}{\partial \tau} = \frac{\partial^2 v}{\partial x^2} + (k-1)\frac{\partial v}{\partial x} - kv \tag{ii}$$

Where $k = r \Big/ \tfrac{1}{2}\sigma^2$. So, the initial condition changes to

$$v(x,0) = \max(e^x - 1, 0)$$

Using the method of change of variable, we construct

$$v = e^{\alpha x + \beta \tau} u(x,\tau),$$

for some constants $\alpha$ and $\beta$, subsequent differentiation yields

$$\frac{\partial v}{\partial \tau} = e^{\alpha x + \beta \tau}\frac{\partial u}{\partial \tau} + \beta e^{\alpha x + \beta \tau} u(x,\tau)$$

$$\frac{\partial v}{\partial \tau} = e^{\alpha x + \beta \tau}\frac{\partial u}{\partial \tau} + \alpha e^{\alpha x + \beta \tau} u(x,\tau)$$

$$\frac{\partial^2 v}{\partial x^2} = e^{\alpha x + \beta \tau}\frac{\partial^2 u}{\partial x^2} + \alpha e^{\alpha x + \beta \tau}\frac{\partial u}{\partial x} + \alpha^2 e^{\alpha x + \beta \tau} u + \alpha e^{\alpha x + \beta \tau}\frac{\partial u}{\partial x}$$

$$\frac{\partial^2 v}{\partial x^2} = \alpha^2 e^{\alpha x + \beta \tau} u + 2\alpha e^{\alpha x + \beta \tau}\frac{\partial u}{\partial x} + e^{\alpha x + \beta \tau}\frac{\partial^2 u}{\partial x^2}$$

$$e^{\alpha x + \beta \tau}\left(\beta u + \frac{\partial u}{\partial \tau}\right) = e^{\alpha x + \beta \tau}\left(\alpha^2 u + 2\alpha\frac{\partial u}{\partial x} + \frac{\partial^2 u}{\partial x^2}\right) + (k-1)e^{\alpha x + \beta \tau}\left(\alpha u + \frac{\partial u}{\partial x}\right) - kue^{\alpha x + \beta \tau}$$

Since, $e^{\alpha x + \beta \tau} \neq 0$

$$\beta u + \frac{\partial u}{\partial \tau} = \alpha^2 u + 2\alpha\frac{\partial u}{\partial x} + \frac{\partial^2 u}{\partial x^2} + (k-1)\left(\alpha u + \frac{\partial u}{\partial x}\right) - ku$$

Choose $\beta = \alpha^2 + (k-1)\alpha - k$ and $0 = 2\alpha + (k-1)$, so that the PDE assumes the standard parabolic form

$$\left(\beta - \alpha^2 - (k-1)\alpha + k\right)u + \frac{\partial u}{\partial \tau} = \alpha^2 u + (2\alpha + (k-1))\frac{\partial u}{\partial x} + \frac{\partial^2 u}{\partial x^2}; -\infty < x < \infty, \tau > 0$$

The argument follows from the fact that $u$ & $\frac{\partial u}{\partial x}$ should vanish and so we have,

$$\alpha = -\frac{1}{2}(k-1), \quad \beta = -\frac{1}{4}(k+1)^2$$

Therefore,

$$v = e^{-\frac{1}{2}(k-1)x - \frac{1}{4}(k+1)^2 \tau} u(x, \tau)$$

Where

$$\frac{\partial u}{\partial \tau} = \frac{\partial^2 u}{\partial x^2}; \; -\infty < x < \infty, \tau > 0,$$

with

$$u(x,0) = u_0(x) = \max\left(e^{\frac{1}{2}(k+1)x} - e^{\frac{1}{2}(k-1)x}, 0\right) \tag{iii}$$

This is, in-fact, the pay-off function for the BS Differential equation.

Lets denote the Fourier transform of $u(x, \tau)$ by

$$u(x,\tau) = \frac{1}{2\sqrt{\pi\tau}} \int_{-\infty}^{\infty} u(x,\tau) e^{i\lambda x^2} dx$$

and assume $u_0(x)$ has a Fourier transform and $u$, $\frac{\partial u}{\partial x}$ vanish at $\infty$ so that the following results might be used

$$\frac{1}{\sqrt{2\pi}} \int_{-\infty}^{\infty} f^n(x) e^{i\lambda x} dx = (-i\lambda)^n F(\lambda); \; n = 1,2,3,........ \tag{iv}$$

Multiplying equation (ii), throughout by $\frac{1}{\sqrt{2\pi}} e^{i\lambda x}$ and integrating w.r.t. $x$ from $-\infty$ to $\infty$ and using equation (iv), we obtain

$$\frac{1}{\sqrt{2\pi}} \int_{-\infty}^{\infty} u_\tau e^{i\lambda x} dx + \lambda^2 u(\lambda,\tau) + \lambda^2 u(\lambda,\tau) + \lambda^2 u(x,t) = 0 \tag{v}$$

and equation (iii) leads to

$$u(\lambda,0) = F(\lambda) = \frac{1}{\sqrt{2\pi}} \int_{-\infty}^{\infty} e^{i\lambda x} u_0(x) dx$$

The solution to the Initial Value Problem is

$$u(\lambda, \tau) = F(\lambda) e^{-\lambda^2 \tau}$$

To find $u(x, \tau)$, apply inverse Fourier Transform and thus we obtain;

$$u(x, \tau) = \frac{1}{\sqrt{2\pi}} \int_{-\infty}^{\infty} e^{-i\lambda x - \lambda^2 \tau} F(\lambda) d\tau$$

$$= \frac{1}{\sqrt{2\pi}} \int_{-\infty}^{\infty} \int_{-\infty}^{\infty} e^{-i\lambda(x-s) - \lambda^2 \tau} u_0(s) d\tau ds \qquad \text{(vi)}$$

Next, evaluate the inner integral,

$$\int_{-\infty}^{\infty} e^{-i\lambda(x-s) - \lambda^2 \tau} d\tau = \int_{-\infty}^{0} e^{-i\lambda(x-s) - \lambda^2 \tau} d\lambda + \int_{0}^{\infty} e^{-i\lambda(x-s) - \lambda^2 \tau} d\lambda$$

$$= 2 \int_{0}^{\infty} e^{-\lambda^2 \tau} \cos[\lambda(x-s)] d\lambda$$

The last integral can be evaluated explicitly. Let the integral $I(\alpha)$ be defined as

$$I(\alpha)\Big|_{\alpha = x-s} = 2 \int_{0}^{\infty} e^{-\lambda^2 \tau} \cos(\alpha \lambda) d\lambda;$$

Note that the integrand is exponentially decaying and hence bounded above.

Differentiating under the integral sign,

$$\frac{dI(\alpha)}{d\alpha} = -2 \int_{0}^{\infty} \lambda e^{-\lambda^2 \tau} \sin(\alpha \lambda) d\lambda - \frac{1}{\tau} \int_{0}^{\infty} \sin(\alpha \lambda) d\left(e^{-\lambda^2 \tau}\right)$$

$$= -\frac{\alpha}{2\tau} I(\alpha)$$

and

$$I(0) = 2 \int_{0}^{\infty} e^{-\lambda^2 \tau} d(\lambda) = \sqrt{\frac{\pi}{\tau}}$$

Combining,

$$I(\alpha)\Big|_{\alpha = x-s} = 2 \int_{0}^{\infty} e^{-\lambda^2 \tau} \cos(\lambda(x-s)) d\lambda$$

$$= \sqrt{\frac{\pi}{\tau}} e^{-(x-s)^2 / 4x} \qquad \text{(vii)}$$

Using equation (vii)

$$u(x,\tau) = \frac{1}{\sqrt{4\pi\tau}} \int_{-\infty}^{\infty} u_0(s) e^{-(x-s)^2/4\tau} ds$$

$$\therefore u(x,\tau) = \frac{1}{2\sqrt{\pi\tau}} \int_{-\infty}^{\infty} u_0(s) e^{-(x-s)^2/4\tau} ds \quad \text{(viii)}$$

where $u_0(x)$ is given by equation (iii). Again using the change of variable,

$$x' = (s-x)/\sqrt{2\tau}$$

We have,

$$u(x,\tau) = \frac{1}{2\sqrt{\pi}} \int_{-\infty}^{\infty} u_0(x'\sqrt{2\tau} + x) e^{-\frac{1}{2}x'^2} dx'$$

$$u(x,\tau) = I_1 - I_2$$

where

$$I_1 = \frac{1}{2\sqrt{\pi}} \int_{-x/\sqrt{2\tau}}^{\infty} e^{\frac{1}{2}(k+1)(x+x'\sqrt{2\tau}) - \frac{1}{2}x'^2} dx'$$

$$\therefore I_1 = e^{\frac{1}{2}(k+1)x + \frac{1}{4}(K+1)^2 \tau} N(d_1)$$

where

$$d_1 = \frac{x}{\sqrt{2\tau}} + \frac{1}{2}(k+1)\sqrt{2\tau}$$

and

$$N(d_1) = \frac{1}{\sqrt{2\pi}} \int_{-\infty}^{d_1} e^{-\frac{1}{2}s^2} ds$$

is the cumulative distribution function for the normal distribution. The calculation of $I_2$ is identical to that of $I_1$ except that $(k+1)$ is replaced by $(k-1)$ throughout.

Now, substituting from equation (i) to recover

$$C(S,t) = SN(d_1) - Ee^{-r(T-t)} N(d_2) \quad \text{(ix)}$$

where

$$d_1 = \frac{\log(S/E) + \left(r + \frac{1}{2}\sigma^2\right)(T-t)}{\sigma(\sqrt{T-t})}$$

$$d_2 = \frac{\log(S/E) + \left(r - \frac{1}{2}\sigma^2\right)(T-t)}{\sigma(\sqrt{T-t})}$$

The corresponding calculation for a European Put Option follows similar lines. Its transformed pay-off is

$$u(x,0) = \max\left(e^{\frac{1}{2}(k-1)z} - e^{\frac{1}{2}(k+1)x}, 0\right) \quad (x)$$

and can be computed as above. However, having evaluated the Call, a simpler way is to use the put-call parity formula

$$C - P = S - Ee^{-r(T-t)}$$

for the value $P$ of a Put, given the value of the Call. This yields

$$P(S,T) = Ee^{-r(T-t)}N(-d_2) - SN(-d_1)$$